# Role of exchange splitting and ligand field splitting in tuning the magnetic anisotropy of an individual iridium atom on TaS$_2$ substrate

Shiming Yan[1] [*], Wen Qiao[1], Deyou Jin[1], Xiaoyong Xu[2], Wenbo Mi[3], Dunhui Wang[1]

*1 School of Electronics and Information, Hangzhou Dianzi University, Hangzhou 310018, China*

*2 School of Physics Science and Technology, Yangzhou University, Yangzhou, 225002, China*

*3 Tianjin Key Laboratory of Low Dimensional Materials Physics and Preparation Technology, School of Science, Tianjin University, Tianjin 300354, China*

## Abstract

In this work, using first-principles calculation we investigate the magnetic anisotropy (MA) of single-atom iridium (Ir) on TaS$_2$ substrate. We find that the strength and direction of MA in the Ir adatom can be tuned by strain. The MA arises from two sources, namely spin-conservation term and spin-flip term. The spin-conservation term is mainly generated by spin-orbit coupling (SOC) interaction on the $d_{xy}/d_{x2-y2}$ orbitals and is contributed to the out-of-plane MA. The spin-flip term is caused by SOC interaction on the $d_{xz}/d_{yz}$ and $p_x/p_y$ orbitals and is responsible for the in-plane MA. We further find that strain-tuned MA is mainly determined by exchange splitting and ligand field splitting. Increase of strain will reduce the ligand field splitting and enhance the exchange splitting, resulting in the enhancement of the out-of-plane MA from $d_{xy}/d_{x2-y2}$ orbitals and the reduction of the in-plane MA from $d_{xz}/d_{yz}$ and $p_x/p_y$ orbitals, and hence leading to the change of the strength and direction of the total MA. Our study provides a way for tuning the MA of single-atoms magnet on 2D transition metal dichalcogenides substrate by control of the exchange splitting and the ligand field splitting.

[*] Corresponding author: shimingyan@hdu.edu.cn

# I. INTRODUCTION

The key to the success of computer and information technology is the continuous miniaturization of solid electronic devices. One of requirement for this improvement is to minimize the unit size of data storage and logic operation. Single-atom magnets (SAMs), which can provide discrete atomic spin states as quantum bits of information for storage and processing, are the ultimate size limit of magnetic information bit [1-4]. The critical feature of the SAMs for realizing the application in information storage and computing is the magnetic anisotropy (MA) [5-12]. Data storage using SAMs requires large magnetic anisotropy energy (MAE) to counteract thermal fluctuation and stabilize the spin bits. At the same time, the energy consumed by switching the magnetization should be minimized in operating the spin bits for information processing. This requires a tunable MA to assist spins manipulation [13-17].

Due to the unique geometric structure and large specific surface areas, two-dimensional (2D) materials are ideal supports as the substrates for single atoms. For the 2D transition metal dichalcogenides (TMDs), the individual atom can be anchored on the surface of basal plane by chemical bonding with the chalcogen atoms. Recently, a number of research groups have successfully fabricated the functional single atoms anchored on the TMDs [17-20]. Because of the low coordination and low symmetry, large orbital moment and spin moment can be induced and enhanced in this system. As a consequence, a large MAE could also be easy generated under the spin-orbit coupling (SOC) interaction. Indeed, many studies have reported the MA of single atom on 2D TMDs substrate. For instance, Odkhuu found a large MAE in the single atom Os on $MoS_2$ [10]; Cong et al. investigated the stabilities and the MA of single transition-metal atoms (TMs) absorbed on the pristine and defective $MoS_2$ monolayer, and found a perpendicular ME with MAE of ~34 meV for Co atoms [6]; Mi et al. studied the MA of 5d TMs adsorbed on monolayer $WSe_2$ and found that the MAE of Ir adatom is ~14 meV [21]; Wu et al. reported a large perpendicular MA (~30 meV) in Re adatoms on $WS_2$ [12].

The SOC interaction is thought to be the main contribution to the MA in magnetic materials. Other interactions, such as magnetic dipole *etc.*, are usually found to play minor roles in inducing the MA [22]. Besides constructing suitable structure configuration, another way to enhance the SOC interaction is to introduce heavy atom, as it has large SOC constant. The enhanced SOC by heavy atom is not only confined on the heavy atom itself but also can extend to the neighboring magnetic atoms due to the proximity effects. Ta element has atomic number of 73 and is usually used as the effect-enhancing ingredient in introducing the SOC in systems. Characteristic example of SOC-related effects enhanced by adding Ta is the ferromagnet/insulator nanojunctions system [23]. In this system, by means of epitaxial growth of Ta metal film as the capping layer, various SOC-related effects can be induced and enhanced, such as spin-orbital touque effect [24-26], spin hall effect [27, 28], Dzyaloshinskii-Moriya interaction effect [29], and large MA effect [15, 23] *etc.* Also because of the large SOC constant of Ta, the intercalated TMs between the layers of 2D tantalum disulfide ($TaS_2$) have large MA [30, 31]. The TMs intercalation compounds can be regarded as the stacking structure of single layers with adsorbed TMs. From this perspective, magnetic single atom adsorbed on the 2D $TaS_2$ monolayer could also have profoundly SOC-related MA effects, and hence the 2D $TaS_2$ might be an ideal substrate for the SAMs.

For common magnetic materials, various approaches have been developed to tune the MAE, such as growing epitaxial film with heavy metals [14, 15, 32], surface charging [33-35], adsorbing molecules on surface [36-38], interface engineering [39-43], applying an electric field [13-17], adding strain [21, 44-46] *etc.* Among these methods, strain engineering was shown to be an effective approach to tune not only the strength but also the easy-axis orientation of the MA [44]. For an individual magnetic atom on 2D TMDs substrate, the strain can be added on the substrate. Although, many reports found remarkable MA for the single magnetic atoms on the 2D TMDs substrate, few researches focused on the influence of substrate strain on the MA and in particular the corresponding mechanism.

In this work, considering the large SOC constant, we chose the Ir atom as the adatom, and the $TaS_2$ as the substrate. Using first-principles calculation we investigate the dependence of MA on the strain for the single-atom Ir on TaS$_2$ substrate. We find that the magnetic moment, orbital moment and MA can be tuned by strain. The origins of the magnetism and MA, and the dependence of strain are explored based on the influences of SOC, exchange splitting and ligand field splitting. We find that strain dependence of MA can be attributed to the changes of the exchange splitting of $d_{xz}/d_{yz}$ and $p_x/p_y$ orbitas and the ligand field splitting of $d_{xy}/d_{x2-y2}$ orbitals. Our study provides a way for tuning the MA in the Ir single-atoms magnetic systems. For example, to enhance the perpendicular MA in this system, we can try to enlarge the exchange splitting and reduce the ligand field splitting.

## II. COMPUTATIONAL DETAILS

All calculations were performed within the framework of density functional theory (DFT) implemented in the Vienna *ab initio* simulation package (vasp). The exchange correlation potential was treated with the generalized gradient approximation (GGA) with the Perdew-Burke-Ernzerh (PBE). The ion-electron interaction was described by the projector-augmented plane-wave (PAW) potentials. All calculations were carried out with a plane-wave energy cut-off of 470 eV and an energy convergence threshold of $10^{-6}$ eV. The Monkhorst-Pack k-point mesh of $7 \times 7 \times 1$ and $21 \times 21 \times 1$ was set for ionic relaxations and noncollinear calculations, respectively. The forces acting on each atom were less than $10^{-2}$ eV/Å. The Ir anchored TaS$_2$ systems were modeled by a supercell of lateral size (4×4). A vacuum region of 20 Å is used to decouple the periodic images.

There are several methods to calculate the MAE: i) self-consistent scheme [17, 42], ii) force theorem [44], iii) torque method [9], iv) Bruno formula [47], and v) four-state method [48]. Here, we used the force theorem approach, in which the SOC interaction is introduced as a perturbation to the scalar relativistic Hamiltonian. The calculation is performed in two steps: i) a first collinear self-consistent field calculation without SOC to obtain the charge density of the ground state of the system; ii) a one-step noncollinear non-self-consistent calculation of two different magnetization directions including SOC. MAE is obtained based on the total energy difference when the magnetization directions are in the $xy$ plane ($E^{\parallel}$) and along the $z$ axis ($E^{\perp}$), MAE = $E^{\parallel}$ − $E^{\perp}$. Positive and negative values of MAE show the easy magnetization axis along the out-of-plane and in-plane directions, respectively. The element and orbital resolved MAE are obtained from the difference between the SOC energies [13, 44] $\Delta E_{soc} = E_{soc\parallel} - E_{soc\perp}$. Here $E_{soc} = \left\langle \frac{h^2}{2m^2c^2} \frac{1}{r} \frac{dV}{dr} \hat{L} \cdot \hat{S} \right\rangle$, where $V(r)$ is the spherical part of the effective potential within the PAW sphere, and $L$ and $S$ are the orbital and spin angular momentums, respectively.

## III. RESULTS AND DISCUSSION

On the basal plane of the TaS$_2$, there are three possible Ir-anchored sites, Ta atop site, S atop site and hollow site, as shown in Fig. 1. The calculated binding energy (defined as $E_b = E_{TaS_2} + E_{Ir} - E_{TaS_2+Ir}$) for Ir anchoring is 5.708 eV, 0.015 eV and 4.527 eV for Ta atop site, S atop site and hollow site, respectively. These values indicate that the Ta atop site is the most stable site for the Ir atom. Such preferred occupation of adatoms on the atop sites of the metal atoms in TMDs substrate is also observed and verified in many previous works [10, 12, 18, 20, 49]. This behavior can be attributed to the additional bonding with the metal atoms other than the S atoms at the top site of the metal atom in TMDs. On the other two sites, only bonding with S atoms appears. [49].

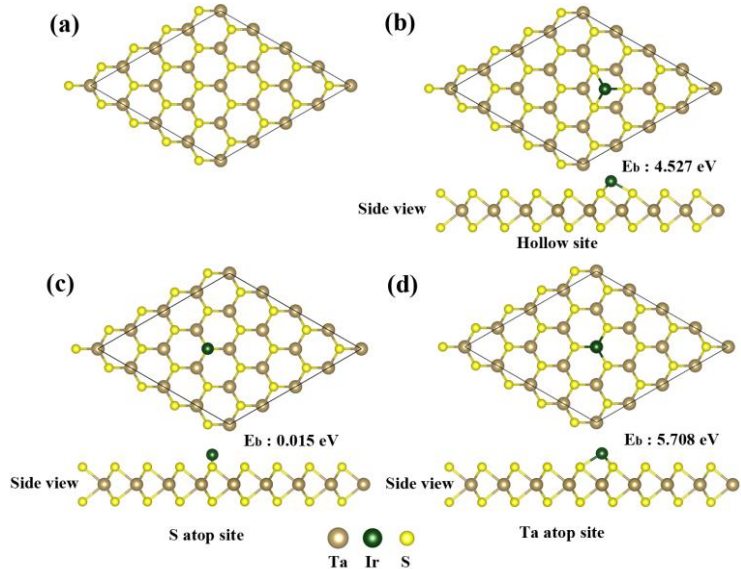

FIG. 1. TaS$_2$ supercell and three possible configuration of Ir anchored on the basal plane of TaS$_2$.

Based on the above calculated results of the binding energy, we select the most stable configuration, Ir anchoring on the Ta atop site, to further investigate the MA. We calculated the MAE with strain and without strain. At free of strain, the calculated total MAE of supercell is ~7.2 meV, suggesting the perpendicular MA in this system. When strain is added, the MAE of Ir atom is notably changed, as shown in Fig. 2(a). As the strain increases from -8% to 8%, the value of MAE of Ir atom increase almost linearly. Tensile strain enhances the out-of-plane MA, while compressive strain results in the reduction of the out-of plane MA and subsequently induces a MA crossover with easy magnetization axis transition from out-of-plane direction to in-plane direction. These results suggest that the strain can efficiency tune the MAE for the single atom Ir on the $TaS_2$.

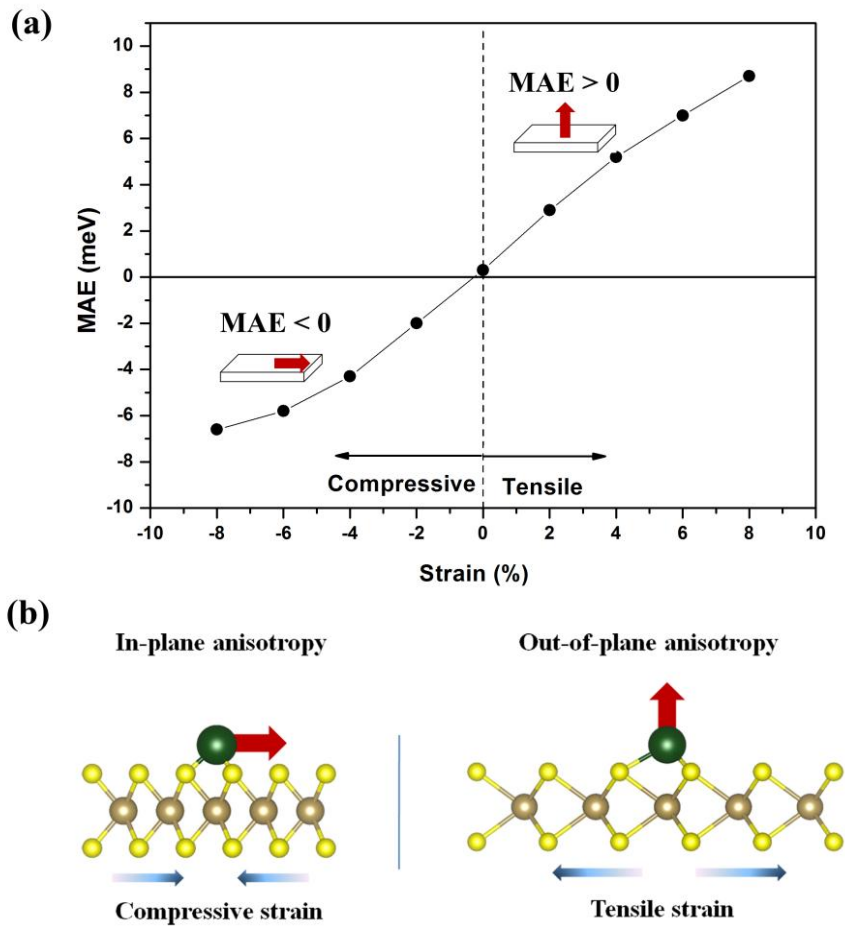

FIG. 2. (a) The MAE of Ir adatom as a function of strain in Ir anchored on $TaS_2$. (b) Schematic diagram of strain dependence of MAE on Ir adatom.

For the single-atom magnet, SOC is the only origin of the MA. Spin magnetic moment and orbital magnetic moment are two basic parameters which are indispensable for the SOC. To understand the influence of strain on the MA, it is necessary to explore spin and orbital moments in this system.

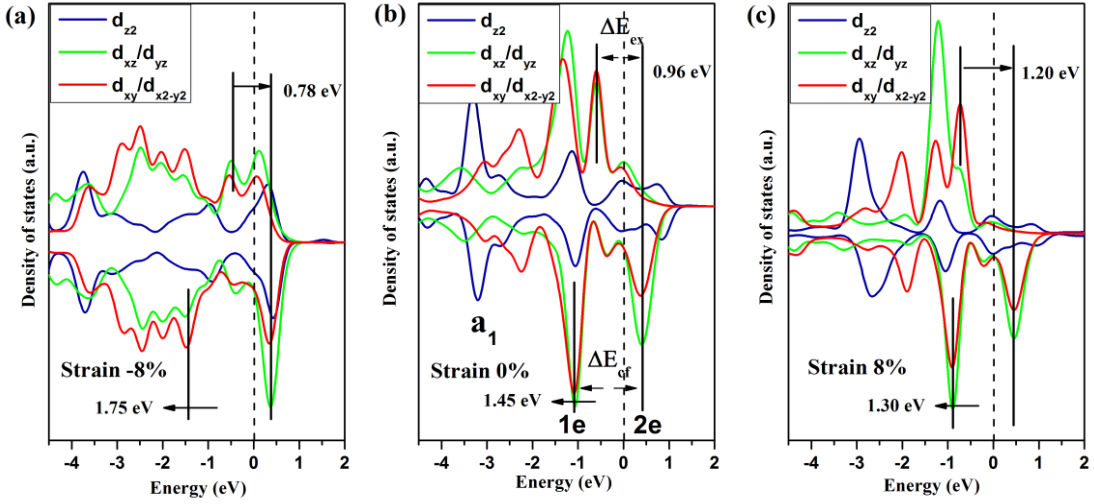

FIG. 3. PDOS of d orbitals of Ir adatom on TaS$_2$ with strain -8%, 0% and 8%.

When there is no strain on the TaS$_2$ substrate, the calculated magnetic moment of Ir adatom is ~0.59 $\mu_B$. From the density of states (DOS) we can analyze the origin of the magnetic moment. As shown in Fig. 3(b), we can see that the $d_{x2-y2}$/ $d_{xy}$ and $d_{xz}$/ $d_{yz}$ orbitals mix with each other and further split to two states (1e and 2e). The coordination environment of Ir on the Ta atop site belongs to the C3v symmetry. Under this ligand field, the $d_{z2}$ orbital forms the basis of a1 irreducible representation and constitutes a single level; the $d_{x2-y2}$/ $d_{xy}$ and $d_{xz}$/$d_{yz}$ orbitals have the same e symmetry, hence they can mix further and produce two double degenerate levels of 1e and 2e [47]. At free of strain, the ligand field splitting energy of 1e and 2e states at Fermi level for Ir adatom is ~ 1.45 eV. The exchange splitting of the e state is ~ 0.96 eV, which is lower than the ligand field splitting. Fig. 4(a) gives the spin-down and spin-up energy levels for d orbital splitting based on the calculated PDOS and the ligand field theory. Considering that there are seven d electrons for Ir atom, spin occupation on the energy levels of d orbitals should be as shown in Fig. 4(a). Accordingly, the magnetic moment for the adsorbed Ir atom should be 1 $\mu_B$. This value is estimated without considering the influence of bonding of d orbitals. Bonding of d orbitals means that d electrons get pairing with other electrons to form spin triplet states and thus the number of unpaired d electrons will reduce, resulting in the decrease of the magnetic moments. Comparing the DOSs of d orbital of Ir adatoms and p orbital of S atoms (As shown in Fig. 4(b)), we can see that parts of the d orbitals indeed participate the bonding with the p orbital of S upon Ir adsorption. Therefore, the calculated magnetic moment of the Ir atom adsorbed on $TaS_2$ is below 1 $\mu_B$.

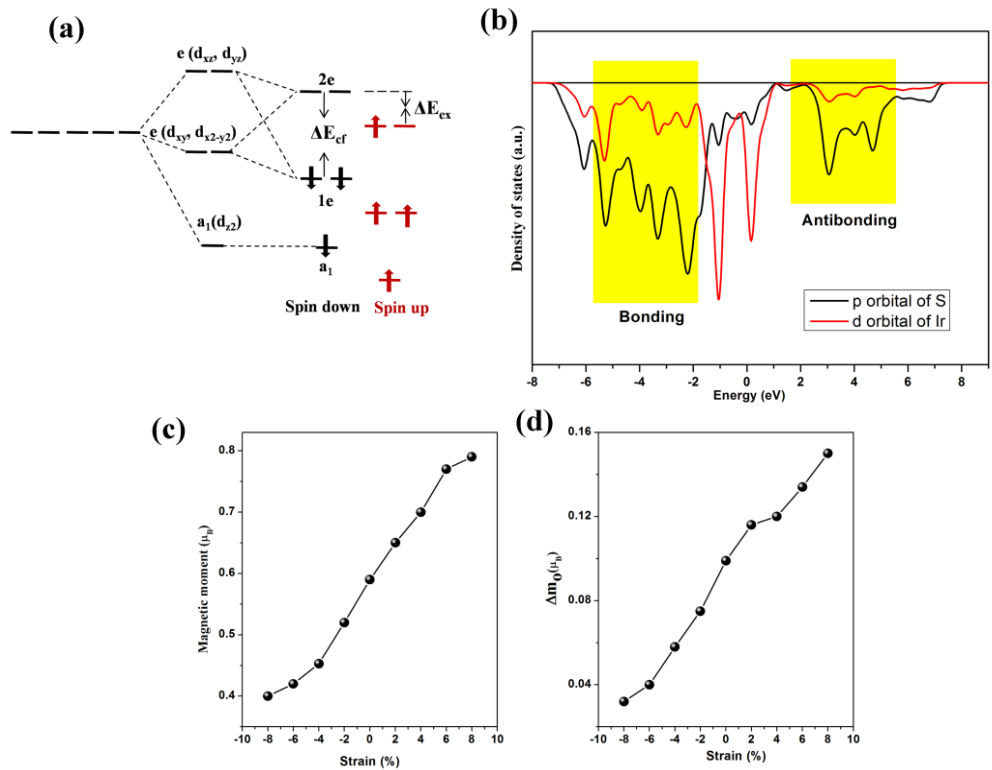

FIG 4. (a) Schematic diagram of energy splitting of d orbital under the influence of ligand field and exchange interaction. (b) DOS of p orbital of S atom and DOS of d orbital of Ir adatom for the Ir-anchored $TaS_2$. (c) and (d) Dependence of strain on the magnetic moment and the difference of orbital moments ($\Delta m_o$).

When strain is added, magnetic moment of Ir is altered notably. As shown in Fig. 4(c), tensile strain enhances the magnetic moment, while compress strain reduces the magnetic moment. One reason is that the bonding strength of d orbitals is changed by strain. Bonding strength not only directly influences the number of unpaired d electrons, but also can influence the exchange splitting of d orbitals. As shown in Fig. 3, tensile strain of 8% gives rise to a large exchange splitting, while compressive strain of -8% results in a small exchange splitting. Large exchange splitting will induce more unpaired d electrons and enhance the magnetic moments. Another reason is the changes of ligand field splitting by strain. Different from the exchange splitting, reducing the tensile strain or enhancing the compressive strain will enlarge the ligand field splitting (as shown in Fig. 3). As the ligand field splitting becomes increasingly large, d electrons in spin-majority states will gradually transfer to spin-minority states, and thus leading to the reduction of magnetic moment.

Fig. 4(d) presents the strain dependence of $\Delta m_o$. Here, the $\Delta m_o$ is the difference of orbital moments along out-of-plane direction and in-plane direction, $\Delta m_o = m_o^\perp - m_o^\parallel$. It can be seen that, the $\Delta m_o$ increases with increasing strain, indicating that lowering the strength of ligand field will enhance the out-of-plane orbital moments. It is noted that the increase of strength of the out-of-plane MA with strain can be explained using the Bruno model. According to the Bruno model, the MAE is proportional to the difference between orbital moments along the easy and hard directions, $MAE = \frac{\xi}{4\mu_B}\Delta m_o$ [50]. As the strain increases, the higher value of $\Delta m_o$, the larger out-of-plane MAE is (Fig. 2(a) and Fig. 4(d)), consistent with the Bruno model. However, the Bruno model could not explain the change of MA direction with strain for the Ir adatom on $TaS_2$. As shown in the Fig. 4(d), within strain range from -8% to 8% the value of $\Delta m_o$ is always positive, suggesting that the out-of-plane orbital moments is invariably dominant. In terms of the Bruno model, the direction of MA should also be out of plane. This is inconsistent with the calculated results of the total MAE of Ir adatom, which shows a transition of direction from out of plane to in plane as strain below 0% (Fig. 2(a)). The reason for this inconsistency is that, as shown below, there is an additional contribution originated from spin-flip term for the MA of Ir adatom. This term is neglected in Bruno model [22, 50, 51].

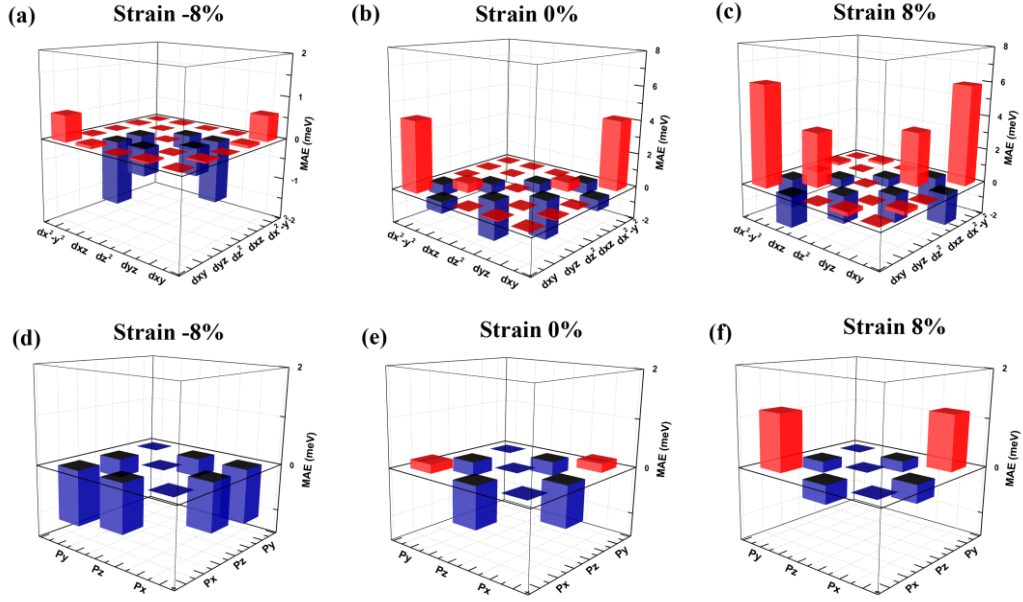

FIG. 5. Orbital-resolved MAE of Ir adatom under strain of -8%, 0% and 8%. (a) (b) (c) are of d orbital; (d) (e) (f) are of p orbital.

The orbital-resolved MAEs of Ir adatom under strain of -8%, 0% and 8% is shown in Fig. 5. One can see that *p* orbitals contribute a part of MAE of Ir adatom. At the strain of -8%, the hybridization of orbitals ($p_x/p_y$, $p_x/p_z$ and $p_y/p_z$) under SOC produce the negative values of MAE, indicating these orbitals are all facilitated to the in-plane MA at this strain. With increasing of strain, the MAE of $p_x/p_y$ orbital changes remarkably from in plane to out of plane. For the *d* orbitals, at strain from -8% to 8%, $d_{xy}/d_{x2-y2}$ orbital contribute to the out-of-plane MA, while $d_{z2}/d_{xz}$ and $d_{z2}/d_{yz}$ orbitals contribute to the in-plane MA. The contribution of $d_{xz}/d_{yz}$ orbital changes from in-plane MA to out-of-plane MA as the strain varies from compression to tension. The MAEs induced by $d_{xy}/d_{x2-y2}$ and $d_{xz}/d_{yz}$ orbitals are sensitive to the strain. But for the $d_{z2}/d_{xz}$ and $d_{z2}/d_{yz}$ orbitals, there are no obvious changes of MAE with altering strain. This is because that the $d_{z2}$ orbital is perpendicular to the basal plane of TaS$_2$ and is insensitive to the in-plane strain.

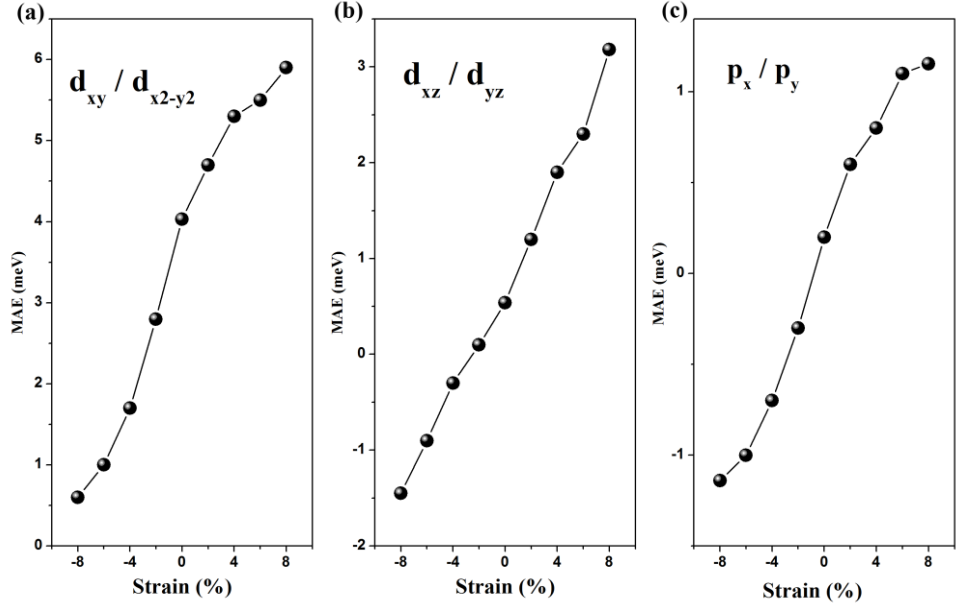

FIG. 6. MAE as a function of strain for the $d_{xy}/d_{x2-y2}$ orbitals, $d_{xz}/d_{yz}$ orbitals and $p_x/p_y$ orbitals.

In order to clearly see the changes of MAE with strain, we plotted the MAE as a function of strain for the strain-sensitive $d_{xy}/d_{x2-y2}$ orbitals, $d_{xz}/d_{yz}$ orbitals and $p_x/p_y$ orbitals (Fig. 6). It can be seen that the positive contribution from the $d_{xy}/d_{x2-y2}$ orbitals increases with increasing strain from -8% to 8%. For the $d_{xz}/d_{yz}$ orbitals and $p_x/p_y$ orbitals, below strain of 0%, they contribute the negative MAE. With increasing strain, the value of the negative MAEs increases and becomes positive at strain above 0%. These monotonic variations with strain induce the change of the direction and strength of MA.

The perturbation theory based on SOC is useful in understanding the mechanism of the influence of orbitals on the MAE. Since the splitting energy of orbitals by ligand field is far larger than the SOC energy, the SOC interaction is regarded as the perturbation on the orbitals is justifiable. Due to the time-reversal symmetry, the real orbitals ($p_x$, $p_y$, $p_z$, $d_{xy}$, $d_{x2-y2}$, $d_{xz}$, $d_{yz}$ and $d_{z2}$) have zero orbital moments, and the SOC energy acted on themselves (one-order perturbation) are also zero. However, the perturbation of SOC on the some mixed orbitals could be nonzero, and thus is responsible for the MAE. Based on the two-order perturbation theory, the MAE come from SOC is expressed by [52]

$$\Delta E^{\downarrow\downarrow} = E^{\downarrow\downarrow}(x) - E^{\downarrow\downarrow}(z)$$

$$= \xi^2 \sum_{o^\downarrow, u^\downarrow} \frac{\left|\left\langle o^\downarrow \left| L_z \right| u^\downarrow \right\rangle\right|^2 - \left|\left\langle o^\downarrow \left| L_x \right| u^\downarrow \right\rangle\right|^2}{E_u^\downarrow - E_o^\downarrow} \quad (1)$$

$$\Delta E^{\uparrow\downarrow} = E^{\uparrow\downarrow}(x) - E^{\uparrow\downarrow}(z)$$

$$= -\xi^2 \sum_{o^\uparrow, u^\downarrow} \frac{\left|\left\langle o^\uparrow \left| L_z \right| u^\downarrow \right\rangle\right|^2 - \left|\left\langle o^\uparrow \left| L_x \right| u^\downarrow \right\rangle\right|^2}{E_u^\downarrow - E_o^\uparrow} \quad (2)$$

Where, u and o are the energy level of the unoccupied states and the occupied states, respectively. ↑ and ↓ are the spin-majority state and spin-minority states, respectively. Here, due to the low proportions of majority unoccupied states near Fermi level (as shown in Fig. 3 and Fig. 7), the Δ$E^{\uparrow\uparrow}$ and Δ$E^{\downarrow\uparrow}$ are neglected. Eq. (1) is the spin-conservation term; Eq. (2) is the spin-flip term. According to the Eq. (1) and (2), the MAE is determined by the spin-orbital matrix element differences as well as their energy differences. Relative contributions of the nonzero $L_z$ and $L_x$ matrix elements are $\left\langle p_x \left| L_z \right| p_y \right\rangle = 1$, $\left\langle d_{xz} \left| L_z \right| d_{yz} \right\rangle = 1$, $\left\langle d_{xy} \left| L_z \right| d_{x^2-y^2} \right\rangle = 2$, $\left\langle d_{z^2} \left| L_x \right| d_{xz}, d_{yz} \right\rangle = \sqrt{3}$, $\left\langle d_{xy} \left| L_x \right| d_{xz}, d_{yz} \right\rangle = 1$ and $\left\langle d_{x^2-y^2} \left| L_x \right| d_{xz}, d_{yz} \right\rangle = 1$ [52].

Fig. 7 gives the PDOS of $p_x/p_y$, $d_{xy}/d_{x2-y2}$ and $d_{xz}/d_{yz}$ orbitals under strain of -8%, 0% and 8%. As the MAE contributed by $d_{xy}/d_{x2-y2}$ orbitals is positive, the MAE should come from the spin-conservation term, $\left\langle d_{xyo}^\downarrow \left| L_z \right| d_{x^2-y^2u}^\downarrow \right\rangle$, which can be viewed as excitations of electrons from the occupied minority-spin states to the unoccupied minority-spin states (as shown in Fig. 7(b)) [22]. For the $d_{xz}/d_{yz}$ and $p_x/p_y$ orbitals, their contributed negative MAE below strain of 0% should both come from the spin-flip term, $\left\langle d_{xzo}^\uparrow \left| L_z \right| d_{yzu}^\downarrow \right\rangle$ and $\left\langle p_{xo}^\uparrow \left| L_z \right| p_{yu}^\downarrow \right\rangle$. Comparing to the Fig. 7(a) and Fig. 7(c), one can see that the $p_x/p_y$ orbitals and the $d_{xz}/d_{yz}$ orbitals exhibit obvious hybridization, and the $p_x/p_y$ orbitals have the same energy splitting as that of $d_{xz}/d_{yz}$ orbitals at Fermi level. Therefore, the $d_{xz}/d_{yz}$ orbitals and $p_x/p_y$ orbitals have the similar contribution for the MAE in Ir adatom.

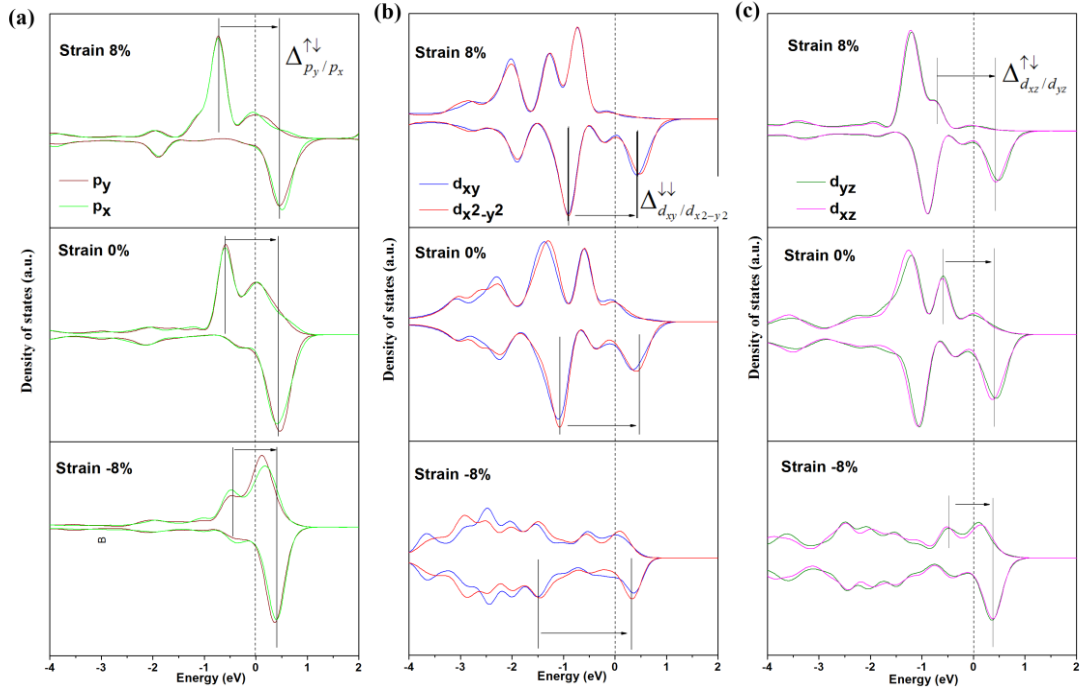

FIG. 7. PDOS of $p_x/p_y$, $d_{xy}/d_{x2-y2}$ and $d_{xz}/d_{yz}$ orbitals of Ir adatom under strain of -8%, 0% and 8%.

Tuning the MAE by strain is mainly through the changes of energy differences of orbitals at Fermi level. Electrons excitations from spin-down 1e state to spin-down 2e state form the spin-conservation term of the $d_{xy}/d_{x2-y2}$ orbitals. The excitations from spin-up 2e state to spin-down 2e state form the spin-flip term of the $d_{xz}/d_{yz}$ and $p_x/p_y$ orbitals. The excited energy is the energy difference, as shown in Fig.7. For the $d_{xz}/d_{yz}$ and $p_x/p_y$ orbitals, the energy difference is their exchange splitting energy at Fermi level. For the $d_{xy}/d_{x2-y2}$ orbitals, the energy difference is the ligand field splitting energy. Small energy difference means easy excitation of electrons, and induces large absolute value of MAE. As shown in Fig. 7, large strain induces big exchange splitting for the $d_{xz}/d_{yz}$ and $p_x/p_y$ orbitals, but reduces the ligand field splitting for the $d_{xy}/d_{x2-y2}$ orbitals. Therefore increase strain will enhance the out-of-plane MA of the $d_{xy}/d_{x2-y2}$ orbitals, but reduce the in-plane MA of the $d_{xz}/d_{yz}$ and $p_x/p_y$ orbitals. Between the range from -8% to 8%, strain can tune the total MAE of Ir adatom monotonously from negative to positive, and thus inducing the transition of MA direction from in plane to out of plane.

The TaS$_2$ substrate plays an important role in tuning the MA of Ir adatom. Essentially, the change of MA with strain in this system results from the modification of the distribution of the non-vertical orbital state of Ir adatom in the electronic structure, which is determined by the TaS$_2$ substrate. As presented above, the strain-tuned MAE of Ir adatom is mainly contributed by the $d_{xz}/d_{yz}$, $p_x/p_y$ and $d_{xy}/d_{x2-y2}$ orbitals because of their appropriate distribution in the electronic structure. The tuning of MAE based on control of SOC is implemented via the strain-induced changes of the exchange splitting of $d_{xz}/d_{yz}$, $p_x/p_y$ orbitals and the ligand field splitting of $d_{xy}/d_{x2-y2}$ orbitals. In contrast to the vertical orbital $d_{z2}$, these orbitals are more sensitive to the in-plane strain, leading to the appearance of the obvious strain effects in the strength and direction of MA in Ir adatom.

As shown in Fig 5, the MAE associated with the vertical $d_{z2}$ orbital is small. This result is also related to the TaS$_2$ substrate. Calculated electronic structure of TaS$_2$ shows that the states of $d_{z2}$ orbital of Ta atom mainly distribute at the Fermi level (Fig S1). When the Ir atom sits on the top site of Ta atom, the $d_{z2}$ orbital of Ir will efficiently hybridize with the $d_{z2}$ orbital of Ta due to the symmetry matching. This hybridization forms σ bonding state below Fermi level and σ$^*$ antibonding state above Fermi level (as shown in Fig 3), and results in the enlargement of bandwidth of $d_{z2}$ state and the decrease of its PDOS at the Fermi level. As a consequence, the MAE related with the $d_{z2}$ orbital, such as $\langle d_{z^2}|L_x|d_{xz}\rangle$ and $\langle d_{z^2}|L_x|d_{yz}\rangle$ terms, is slight for contributing to the total MAE of Ir.

To confirm the effects of substrate on the MA of adsorbed Ir atom, we replaced the TaS$_2$ with MoS$_2$ substrate and calculated the corresponding MAE and PDOS of Ir. We find that, different from the Ir adsorbed on TaS$_2$, the states of $d_{z2}$ orbital of Ir on MoS$_2$ mainly distribute at the Fermi level (Fig S2(b)), and the MAE related with the $d_{z2}$ orbital dominates the total MAE of Ir (Fig S2(a)). For MoS$_2$, there is a band gap in electronic structure, and no states of $d_{z2}$ orbital distribute at the Fermi level. When the Ir atom adsorbs on the MoS$_2$, the states of $d_{z2}$ orbital of the adsorbed Ir can locate at the Fermi level, and giving rise to the large in-plane MA through $\langle d_{z^2}^{\downarrow}|L_x|d_{xz}^{\downarrow}\rangle$ term (as shown in Fig S2). This result is distinctly different from the case of TaS$_2$, where the d$_{z2}$ orbital of Ir gives minor contribution to the total MAE. Therefore, the MA should be the results of the interaction between the substrate and the adsorbed Ir. Substrate play an important role in tuning the MA of the magnetic adatom. For the Ir adatom on $TaS_2$ substrate, the MA of the Ir adatom can be efficiently modified by tuning the substrate.

## IV. CONCLUSION

In summary, we investigated the MA of the single-atom Ir on the $TaS_2$ substrate. We find that the direction and strength of MA in Ir atom can be tuned by strain. The strain-sensitive MAE of Ir can be divided into two parts. One is the negative contribution from the $d_{xz}/d_{yz}$ orbitals and $p_x/p_y$ orbitals. Another is the positive contribution from the $d_{xy}/d_{x2-y2}$ orbitals. Tuning the MAE by strain is mainly through changes of exchange splitting and ligand field splitting. Increase of strain will enhance the exchange splitting and reduce the ligand field splitting, resulting in the enhancement of the in-plane MA and the reduction of the out-of-plane MA, and hence leading to the change of the direction and strength of MA on Ir adatom.

## ACKNOWLEDGEMENTS

This work was supported by the National Natural Science Foundation of China (Grant No. 11504086), the Ten Thousand Talents Plan of Zhejiang Province of China (Grant No. 2019R52014), the Open Project of National Laboratory of Solid State Microstructures, Nanjing University (Grant No. M33010) and the School Scientific Research Project of Hangzhou Dianzi University (Grant Nos. KYS045619084, KYS045619085).